\documentclass[fleqn,twoside]{article}
\usepackage{espcrc2}

\usepackage{graphicx}
\usepackage[figuresright]{rotating}

\newcommand{\AmS}{{\protect\the\textfont2
  A\kern-.1667em\lower.5ex\hbox{M}\kern-.125emS}}

\title{Neutrino Cross Sections at HERA and Beyond}

\author{Mary Hall Reno\address{
        Department of Physics and Astronomy,
        University of Iowa, Iowa City, Iowa 52242
        USA}\thanks{To appear in the Proceedings of the XIII International Symposium
        on Very High Energy Cosmic Ray Interactions, 6-12 September 2004, Pylos, Greece.} }
\begin{document}

\begin{abstract}
The ultrahigh energy neutrino cross section is reviewed. Experimental
results from HERA $ep$ and $\bar{e}p$ scattering have
yielded new information about the small $x$ behavior of the parton
distribution functions. We compare pre-HERA era neutrino-nucleon cross sections
with more recent evaluations.
\vspace{1pc}
\end{abstract}

\maketitle

\section{INTRODUCTION}

The sources of very high energy cosmic rays should also be sources
of neutrinos. While the complete scheme of ultrahigh energy cosmic
ray production is not yet understood \cite{cr},
ancillary to cosmic ray production are neutrino fluxes, both at
the source of the cosmic rays and through cosmic ray propagation
and scattering with the microwave background. Given that neutrinos
are unabsorbed and undeflected between their sources and the Earth,
neutrino telescopes may yield results that allow for a better
understanding of cosmic ray production \cite{lm}. The neutrino-nucleon
cross section is a key ingredient in neutrino telescope measurements.

The theoretical predictions of ultrahigh energy neutrino cross sections
have been greatly influenced by the deep inelastic scattering (DIS)
results of HERA experiments, starting in 1992. 
Neutrino cross sections were not measured at HERA, where beams of
electrons and positrons are incident on proton beams, however, our
knowledge
of parton distribution functions has been dramatically improved
with HERA results. At an energy equivalent to a neutrino energy of 54 TeV, 
the HERA results have aided our understanding of the parton distribution functions
in new kinematic regimes and have led to different high energy extrapolations
of the ultrahigh energy neutrino cross section than what was done earlier.

We review the evaluation of the neutrino-nucleon cross section
and summarize the consequences of HERA results for
the small $x$ region of the parton distribution functions, and therefore, for the
ultrahigh energy neutrino cross section. Alternatives to straightforward QCD evolution of
the parton distribution functions are reviewed and their consequences for the 
ultrahigh energy cross section are compared.

\section{NEUTRINO CROSS SECTION}

We begin with a review of the cross section for
\begin{equation}
\nu_\mu (k)\, N(p)\, \rightarrow \mu (k^\prime)\, X \ ,
\end{equation}
neutrino scattering with an isoscalar nucleon $N$. The charged current
differential cross section, in terms of
$x=Q^2/(2 p\cdot q)$, $Q^2=-q^2$,
$q=k-k^\prime$, $y=p\cdot q/(p\cdot k)$,  and the nucleon mass $M$, 
is 
\begin{eqnarray}
\label{eqn:dsdxdy} 
\frac{d^2\sigma}{dx\ dy} &&= \frac{2G_F^2 M
E_{\nu}}{\pi(1+Q^2/M_W^2)^2}\nonumber \\
&& \times  \left\{
q(x,Q^2)+(1-y)^2\bar{q}(x,Q^2)\right\} \ .
\end{eqnarray} 
The parton distribution functions (PDFs) for the relevant quark and
antiquark distributions are functions of $(x,Q^2$) and labeled as
$q(x,Q^2)$ and $\bar{q}(x,Q^2)$. 

The structure of Eq. (\ref{eqn:dsdxdy}) together with a knowledge of
the Dokshitzer-Gribov-Lipatov-Altarelli-Parisi (DGLAP) evolution \cite{dglap} of the
PDFs give a qualitative understanding of the energy behavior of the
cross section. At low energies, where $Q^2\ll M_W^2$, the PDFs are nearly
$Q^2$ independent, and the neutrino-nucleon cross section scales with
incident neutrino energy $E_\nu$. At high energies, the increase in
the PDFs with increasing $Q^2$ is more than offset by the decrease in
the cross section due to the $W$-boson propagator. As a result, using 
the relation that
\begin{equation}
xy(2ME_\nu)=Q^2\ ,
\end{equation}
and the approximation that $Q^2_{max}\sim M_W^2$, one is led to the
relation 
\begin{equation}
x\sim \frac{10^4}{(E_\nu/{\rm GeV})}\ .
\end{equation}
At the highest energies of interest, on the order of $E_\nu\sim 10^{12}$ GeV,
this translates to $x\sim 10^{-8}$. HERA measurements made a significant improvement
in the range of $(x,Q^2)$ compared to fixed target measurements: at HERA energies,
given that $y\leq 1$, $x\ge 10^{-5}\cdot Q^2$ is accessible. 
Information about small $x$ PDFs at values of $Q^2\sim 1$ GeV$^2$ is then
evolved to higher $Q^2$. For
$Q^2\sim M_W^2$, one has only a limited range in $x$ directly measured at HERA \cite{h1,zeus}.
The kinematic reach in $(x,Q^2)$ at HERA is bigger than for
fixed target DIS, where e.g., for $E=350$ GeV, $x>1.6\times 10^{-3} \cdot Q^2$.
In addition to DIS at HERA and in fixed target experiments, there are PDF measurements
at colliders. Tevatron experimental results are limited to $x>10^{-3}$, however, these measurements
are done at large $Q^2$ values \cite{tevatron}.

\begin{figure}[htb]
%\vspace{9pt}
%\framebox[55mm]{\rule[-21mm]{0mm}{43mm}}
\includegraphics[angle=270,width=17pc]{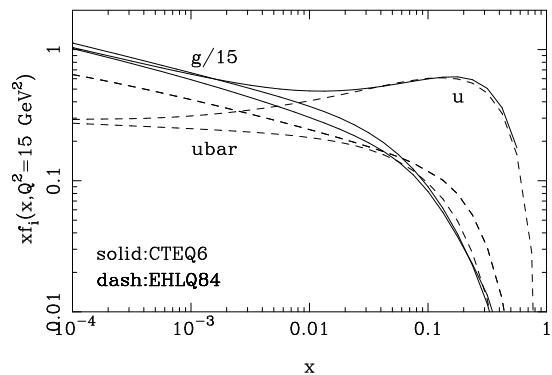}
\caption{Up and $\bar{u}$ PDFs, and the gluon PDF rescaled by a factor of 15, as a 
function of $x$ for $Q^2=15$ GeV$^2$ for the EHLQ PDFs (dash) and CTEQ6 PDFs (solid).}
\label{fig:pdf3p9}
\end{figure}

\begin{figure}[htb]
%\vspace{9pt}
%\framebox[55mm]{\rule[-21mm]{0mm}{43mm}}
\includegraphics[angle=270,width=17pc]{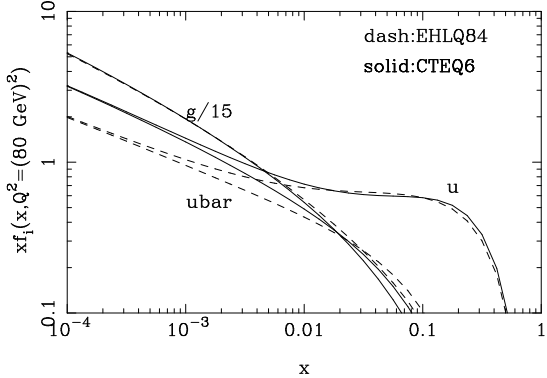}
\caption{As in Fig. 1, now with $Q^2=M_W^2$.}
\label{fig:pdf80}
\end{figure}

Parton distribution function parametrizations are  extracted from 
fits to the world data from DIS and collider experiments. 
Details on the theory and experimental aspects of the extraction of PDFs can be obtained
in Ref. \cite{devcoop}, for example. A number of publications
have provided PDF parametrizations, most recently in Refs. \cite{cteq6,mrst,grv},
with prior versions which have evolved as more data were published.
An interesting comparison is between the parton distribution functions which were
one of the standards in the study of the physics of the proposed Superconducting
Super Collider (SSC) by Eichten, Hinchliffe, Lane and Quigg (EHLQ) from 1984 \cite{ehlq}, with
a current PDF set such as the Pumplin et al.'s CTEQ6 fit \cite{cteq6}. The $u$, $\bar{u}$ and
gluon PDFs are shown in Figs. \ref{fig:pdf3p9}
and \ref{fig:pdf80}  for $Q^2=15$ GeV$^2$ and for
$Q^2=M_W^2$. 

An important feature of the earlier fits was that at the reference value of $Q=Q_0$ at which
the PDFs were parametrized, it was assumed that the sea distribution functions and gluon
distribution function had small $x$ extrapolations following
\begin{equation}
xf(x,Q^2)\sim {\rm constant}\ .
\end{equation}
This is in comparison to a steeper distribution at small $x$ for the modern set, where
in Fig. \ref{fig:pdf3p9} one sees a small $x$ behavior with
\begin{equation}
xf(x,Q^2)\sim x^{-\lambda}\ .
\end{equation}
The small $x$ behavior at lower values of $Q$ has an influence on the extrapolation
to even smaller $x$ values. It is to this subject that we now turn.

\section{SMALL $x$ EXTRAPOLATIONS}

The small $x$ extrapolations of the PDFs in the DGLAP formalism can be inferred from
the behavior of the gluon distribution. Small $x$ PDFs are important at ultrahigh
energies where sea quarks dominate. This is manifest in Fig. \ref{fig:sig} where
the $\nu N$ and $\bar{\nu}N$ charged current cross sections are equal at $E\sim 
10^6 $ GeV. The sea distributions come from gluon splitting to quark-antiquark pairs,
\begin{equation}
g\rightarrow q\bar{q}\ ,
\end{equation}
so the gluon extrapolation at small $x$ is a good guide to the quark sea PDF extrapolations.
For the typical values of $\lambda$ in modern parametrizations, if 
\begin{equation}
xg(x,Q_0^2)\sim A(Q_0^2)\, x^{-\lambda}\ \quad\quad x \ll 1\ ,
\end{equation}
the gluon distribution function has a similar form at higher values of $Q^2$ \cite{ekl}:
\begin{equation}
xg(x,Q^2)\sim A(Q^2)\, x^{-\lambda}\ .
\end{equation}
Our approach \cite{cs1} has been to extrapolate the sea quarks in a similar way,
matching the CTEQ6 PDFs below $x_{min}=10^{-6}$ via
\begin{equation}
x{q}(x,Q^2)=\Biggl( \frac{x_{min}}{x}\Biggr)^{\lambda}
x{q}(x_{min},Q^2)\ .
\end{equation}
The quantity $\lambda$ is determined for each flavor from the PDFs at $Q^2=M_W^2$.
With the earlier EHLQ parametrizations, one would instead be lead to a small $x$
PDF extrapolation in the double-logarithmic-approximation (DLA) \cite{glr}. 

\section{CROSS SECTIONS}

\begin{figure}[htb]
%\vspace{9pt}
%\framebox[55mm]{\rule[-21mm]{0mm}{43mm}}
\includegraphics[angle=270,width=15pc]{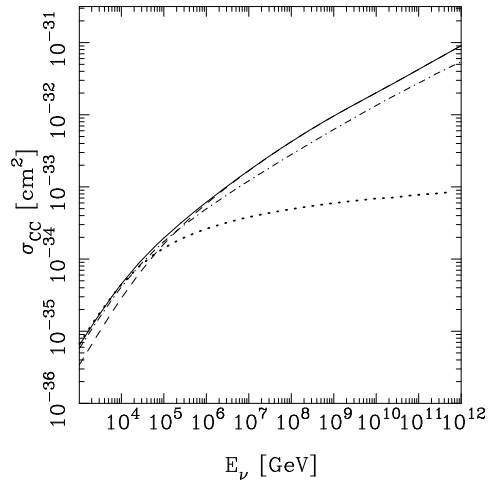}
\caption{The charged current $\nu N$ (solid) and $\bar{\nu} N$ (dashed)
cross sections as a function of incident neutrino/antineutrino energy,
using the CTEQ6 PDFs. Also shown
is the $\nu N$ cross section using the EHLQ PDFs (dot-dashed) and EHLQ PDFs
frozen at the scale of $Q^2=5$ GeV$^2$ (dotted).}
\label{fig:sig}
\end{figure}

It is the cross section evaluated using the
EHLQ PDFs with a DLA extrapolation \cite{cs3} 
that is shown in Fig. \ref{fig:sig} by the dot-dashed
line, while the dashed (for incident $\bar{\nu}$'s) and solid (for incident $\nu$'s) lines
show the results using the power law extrapolation with $\lambda$. Similar results are
obtained using the Gluck, Reya and Vogt \cite{grv} PDFs in the evaluation of the
cross section \cite{cs2}.
For reference,
we also show the charged current cross section using the EHLQ PDFs with $Q^2$ frozen
at $Q_0^2$. 

An important reference point in Fig. \ref{fig:sig} is at $E_\nu\sim 40$ TeV,
at which the neutrino interaction length is approximately equal to the diameter of the
Earth. At higher energies, neutrinos traversing the Earth over trajectories through
the Earth's core are significantly attenuated.

\begin{figure}[htb]
%\vspace{9pt}
%\framebox[55mm]{\rule[-21mm]{0mm}{43mm}}
\includegraphics[angle=270,width=15pc]{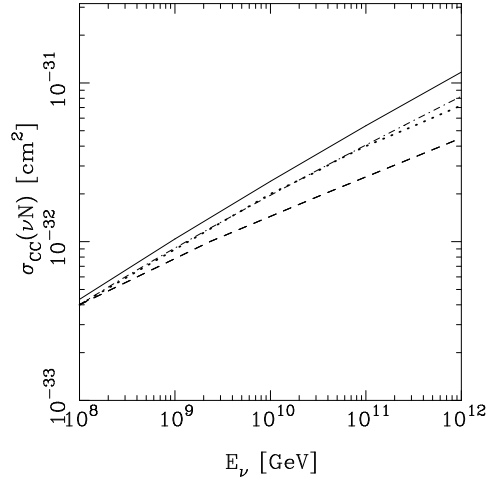}
\caption{The charged current $\nu N$ cross section adapted from
Ref. [21] of Kutak and Kwiecinski, showing $\sigma$ using
a unified BFKL/DGLAP treatment with (dotted) and without (solid) screening.
The CTEQ6 (dot-dash) result and a calculation using the GBW saturation model (dashed)
from Ref. [21] are also shown.}
\label{fig:sigx}
\end{figure}

The importance of QCD evolution in the context of neutrino scattering
was first pointed out by Andreev, Berezinsky and Smirnov in Ref. \cite{abs}. The curves
in Fig. \ref{fig:sig} make it clear that any additional modifications to the
small $x$ PDFs must incorporate $Q^2$ evolution. One such modification is the inclusion
of $\ln(1/x)$ corrections using the Balitsky, Fadin, Kuraev and Lipatov
(BFKL) formalism \cite{bfkl}. Only a unified BFKL/DGLAP approach can be used for
ultrahigh energy neutrino interactions. Kwiecinski, Martin and Stasto (KMS) have made such an
evaluation \cite{kms} of the neutrino cross section, which is shown by the solid line in Fig.
\ref{fig:sigx}. Other theoretical work on the topic of $\ln (1/x)$ corrections
\cite{altarelli} cannot be applied
directly to the neutrino nucleon cross sections  until one has
a form that explicitly includes DGLAP evolution.

Additional corrections have been made by Kutak and Kwiecinski (KK) in Ref. \cite{kk}, 
where a nonlinear term is included in the evolution equations to account for
gluon recombination effects. A continued growth of the small $x$ PDFs with decreasing
$x$ ultimately will conflict with unitarity requirements. In parton language, the
small $x$ PDFs will moderate due to gluon recombination:
\begin{equation}
gg\rightarrow g\ .
\end{equation}
The KK result including recombination effects is shown in Fig. \ref{fig:sigx}
by the dotted line. A different model for screening, the color dipole model of 
Golec-Biernat and Wusthoff (GBW) \cite{gbw}, as calculated
in Ref. \cite{kk} is shown in the same figure with a dashed
line. For reference, we also show the CTEQ6 result using the power law small $x$ extrapolation.
All of the cross sections in Fig. \ref{fig:sigx} are in remarkably good agreement at
$E_\nu=10^{12}$ GeV.

Other references \cite{fiore,machado} have recently appeared, including saturation
or recombination effects. They are consistent with the range of results shown in 
Fig. \ref{fig:sigx}. 

\section{FINAL REMARKS}

Our knowledge of the parton distribution functions has greatly increased
since HERA data became available in the early 1990's. Extrapolations based on DGLAP or a unified
BFKL/DGLAP with and without recombination effects 
to $x$ values far below the measured regime at $Q^2\sim M_W^2$ all yield similar
cross sections at $E_\nu=10^{12}$ GeV.  There are prospects for new measurements of the
PDFs with the upcoming Large Hadron Collider experiments, however, the smallest $x$
values relevant to ultrahigh energy neutrinos are outside our reach in collider experiments.

The small $x$ behavior of the PDFs has implications in
neutrino astrophysics for more than the neutrino-nucleon
cross section. At lower energies, the small $x$ PDFs are inputs to $c\bar{c}$ and
$b\bar{b}$ production by cosmic ray interactions with air nuclei. The semileptonic
heavy quark decays are responsible for the prompt lepton fluxes which should dominate
the atmospheric flux in the range of 100 TeV \cite{martin,earlier}. Measurements
of the prompt lepton flux may help constrain small $x$ physics.

\section*{Acknowledgments}

The author thanks I. Sarcevic and I. Mocioiu for recent discussions and collaboration on the topic
of the ultrahigh energy neutrino cross section, and I. Sarcevic, R. Gandhi and C. Quigg
for their collaboration in years past. The work here was supported in part by  U.S. 
D.O.E. contract DE-FG02-91ER40664.


\begin{thebibliography}{99}

\bibitem{cr}
See, e.g., R. J. Protheroe and R. W. Clay, Publ. Astron. Soc. Pac. 21 (2004) 1;
P. Bhattacharjee and G. Sigl, Phys. Rept. 327 (2000) 109 and references therein.

\bibitem{lm}
J. G. Learned and K. Mannheim, Ann. Rev. Nucl. Part. Sci. 50 (2000) 679.

\bibitem{dglap}
G. Altarelli and G. Parisi, Nucl Phys. B 126 (1977) 298;
L. N. Lipatov, Sov. J. Nucl. Phys. 20 (1975) 95;
V. N. Gribov and L. N. Lipatov, Sov. J. Nucl. Phys. 15 (1972) 438;
Yu. L. Kokshitzer, Sov. Phys. JETP 46 (1977) 641.
See also, e.g., G. Altarelli, Phys. Rep. 81 (1983).

\bibitem{h1}
C.~Adloff {\it et al.}  [H1 Collaboration],
%``Measurement and QCD analysis of neutral and charged current cross  sections
%at HERA,''
Eur.\ Phys.\ J.\ C { 30}(2003) 1.
%[arXiv:hep-ex/0304003].
%%CITATION = HEP-EX 0304003;%%

%\cite{Chekanov:2003vw}
\bibitem{zeus}
S.~Chekanov {\it et al.}  [ZEUS Collaboration],
%``Measurement of high-Q**2 charged current cross sections in e+ p deep
%inelastic scattering at HERA,''
Eur.\ Phys.\ J.\ C { 32} (2003) 1;
%%CITATION = HEP-EX 0307043;%%
%\cite{Chekanov:2003yv}
%\bibitem{Chekanov:2003yv}
S.~Chekanov {\it et al.}  [ZEUS Collaboration],
%``High-Q**2 neutral current cross sections in e+ p deep 
%inelastic scattering at
%s**(1/2) = 318-GeV,''
arXiv:hep-ex/0401003.
%%CITATION = HEP-EX 0401003;%%


\bibitem{tevatron}
See, e.g., Refs. [7-9] and references therein.

\bibitem{devcoop}
R. Devenish and A. Cooper-Sarkar, {\it Deep Inelastic
Scattering} (Oxford University Press, Oxford, 2004).


\bibitem{cteq6}
J.~Pumplin, D.~R.~Stump, J.~Huston, H.~L.~Lai, P.~Nadolsky and W.~K.~Tung,
%``New generation of parton distributions with uncertainties from global  QCD
%analysis,''
JHEP { 0207} (2002) 012.
%[arXiv:hep-ph/0201195].
%%CITATION = HEP-PH 0201195;%%

\bibitem{mrst}
A. D. Martin, R. G. Roberts, W. J. Stirling and R. S. Thorne,
Eur. Phys. J. C 28 (2003) 455; Eur. Phys. J. C 35 (2004) 325.

\bibitem{grv}
M.~Gluck, E.~Reya and A.~Vogt,
%``Dynamical parton distributions revisited,''
Eur.\ Phys.\ J.\ C { 5} (1998) 461.
%[arXiv:hep-ph/9806404].
%%CITATION = HEP-PH 9806404;%%


\bibitem{ehlq}
E.~Eichten, I.~Hinchliffe, K.~D.~Lane and C.~Quigg,
%``Super Collider Physics,''
Rev.\ Mod.\ Phys.\  {56} (1984) 579
[Addendum-ibid.\  { 58} (1986) 1065].
%%CITATION = RMPHA,56,579;%%
See also, D.~W.~Duke and J.~F.~Owens,
%``Q**2 Dependent Parametrizations Of Parton Distribution Functions,''
Phys.\ Rev.\ D { 30} (1984) 49.
%%CITATION = PHRVA,D30,49;%%

\bibitem{ekl}
R. K. Ellis, Z. Kunszt and E. M. Levin,
Nucl. Phys. B 420 (1994) 517.

\bibitem{cs1}
R.~Gandhi, C.~Quigg, M.~H.~Reno and I.~Sarcevic,
%``Neutrino interactions at ultrahigh energies,''
Phys.\ Rev.\ D { 58} (1998) 093009.
%[arXiv:hep-ph/9807264].
%%CITATION = HEP-PH 9807264;%%
See also 
R.~Gandhi, C.~Quigg, M.~H.~Reno and I.~Sarcevic,
%``Ultrahigh-energy neutrino interactions,''
Astropart.\ Phys.\  {5}(1996) 81;
%%CITATION = HEP-PH 9512364;%%
G. M. Frichter, D. W. McKay and J. P. Ralston,
Phys. Rev. Lett. 74  (1995) 1508.

\bibitem{glr}
L. V. Gribov, E. M. Levin and M. G. Ryskin,
Phys. Rep. 100 (1983) 1.
%\cite{Gandhi:1995tf}

\bibitem{cs3}
Earlier evaluations of the cross section, before the results of
HERA experiments, appear in, e.g.,
D. W. McKay and J. P. Ralston,
Phys. Lett. 167B (1986) 103;
C.~Quigg, M.~H.~Reno and T.~P.~Walker,
%``Interactions Of Ultrahigh-Energy Neutrinos,''
Phys.\ Rev.\ Lett.\  { 57} ( 1986) 774.
%%CITATION = PRLTA,57,774;%%


\bibitem{cs2}
M.~Gluck, S.~Kretzer and E.~Reya,
%``Dynamical QCD predictions for ultrahigh energy neutrino cross sections,''
Astropart.\ Phys.\  {11} (1999) 327.
%[arXiv:astro-ph/9809273].
%%CITATION = ASTRO-PH 9809273;%%

\bibitem{abs}
Yu. M. Andreev, V. S. Berezinsky and A. Yu. Smirnov, Phys.
Lett. B 143 (1978) 521.



\bibitem{bfkl}
E.~A.~Kuraev, L.~N.~Lipatov and V.~S.~Fadin,
%``The Pomeranchuk Singularity In Nonabelian Gauge Theories,''
Sov.\ Phys.\ JETP {45} (1977) 199
[Zh.\ Eksp.\ Teor.\ Fiz.\  {72} (1977) 377];
%%CITATION = SPHJA,45,199;%%
I.~I.~Balitsky and L.~N.~Lipatov,
%``The Pomeranchuk Singularity In Quantum Chromodynamics,''
Sov.\ J.\ Nucl.\ Phys.\  { 28} (1978) 822
[Yad.\ Fiz.\  { 28} (1978) 1597].
%%CITATION = SJNCA,28,822;%%

\bibitem{kms}
J. Kwiecinski, A. Martin and A. Stasto,
Phys. Rev. D 56 (1997) 3991.

\bibitem{altarelli}
G. Altarelli, R. D. Ball and S. Forte,
Nucl. Phys. B 674 (2003) 459;
M.~Ciafaloni, D.~Colferai, G.~P.~Salam and A.~M.~Stasto,
%``Renormalisation group improved small-x Green's function,''
Phys.\ Rev.\ D {68} (2003) 114003;
%[arXiv:hep-ph/0307188].
%%CITATION = HEP-PH 0307188;%%
M. Ciafaloni, D. Colferai, G. P. Salam and
A. M. Stasto,
Phys. Lett. B 587 (2004) 87;
R. S. Thorne, Phys. Rev. D 64 (2001) 074005.


%\cite{kk}
\bibitem{kk}
K.~Kutak and J.~Kwiecinski,
%``Screening effects in the ultrahigh energy neutrino interactions,''
Eur.\ Phys.\ J.\ C {29} (2003) 521. 
%[arXiv:hep-ph/0303209].
%%CITATION = HEP-PH 0303209;%%

%\cite{Golec-Biernat:1998js}
\bibitem{gbw}
K.~Golec-Biernat and M.~Wusthoff,
%``Saturation effects in deep inelastic scattering at low Q**2 and its
%implications on diffraction,''
Phys.\ Rev.\ D { 59} (1999) 014017.
%[arXiv:hep-ph/9807513].
%%CITATION = HEP-PH 9807513;%%

%\cite{Fiore:2003kc}
\bibitem{fiore}
R.~Fiore, L.~L.~Jenkovszky, A.~Kotikov, F.~Paccanoni, A.~Papa and E.~Predazzi,
%``Ultra-high energy neutrino nucleon interactions,''
Phys.\ Rev.\ D {68} (2003) 093010.
%[arXiv:hep-ph/0302251].
%%CITATION = HEP-PH 0302251;%%

\bibitem{machado}
M.~V.~T.~Machado,
%``Ultrahigh energy neutrinos and non-linear QCD dynamics,''
Phys. Rev. D 70 (2004) 053008.
%%CITATION = HEP-PH 0311281;%%



\bibitem{martin}
A.~D.~Martin, M.~G.~Ryskin and A.~M.~Stasto,
%``Prompt neutrinos from atmospheric c anti-c and b anti-b production and  the
%gluon at very small x,''
Acta Phys.\ Polon.\ B {34} (2003) 3273.
%[arXiv:hep-ph/0302140].
%%CITATION = HEP-PH 0302140;%%

\bibitem{earlier}
For earlier work on prompt neutrinos and the gluon distribution
function, see e.g.,
G.~Gelmini, P.~Gondolo and G.~Varieschi,
%``Prompt atmospheric neutrinos and muons: Dependence on the gluon
%distribution function,''
Phys.\ Rev.\ D {61} (2000) 056011;
%[arXiv:hep-ph/9905377].
%%CITATION = HEP-PH 9905377;%%
G.~Gelmini, P.~Gondolo and G.~Varieschi,
%``Measurement of the gluon PDF at small x with neutrino telescopes,''
Phys.\ Rev.\ D {63} (2001) 036006;
%[arXiv:hep-ph/0003307].
%%CITATION = HEP-PH 0003307;%%
L.~Pasquali, M.~H.~Reno and I.~Sarcevic,
%``Lepton fluxes from atmospheric charm,''
Phys.\ Rev.\ D {59} (1999) 034020.
%[arXiv:hep-ph/9806428].
\end{thebibliography}
\end{document}